\documentclass[preprint,showpacs,preprintnumbers,amsmath,amssymb,nofootinbib]{revtex4}

\usepackage{booktabs}
\usepackage{mathrsfs}
\usepackage{epsfig}
\usepackage{graphicx}
\usepackage{dcolumn}
\usepackage{bm}
\usepackage{amsmath}
\usepackage{slashed}
\usepackage{multirow}
\usepackage{subfigure}
\usepackage{epstopdf}
\usepackage{color}

\let\jnfont=\rm
\def\NPB#1,{{\jnfont Nucl.\ Phys.\ B }{\bf #1},}
\def\PLB#1,{{\jnfont Phys.\ Lett.\ B }{\bf #1},}
\def\EPJC#1,{{\jnfont Eur.\ Phys.\ Jour.\ C }{\bf #1},}
\def\PRD#1,{{\jnfont Phys.\ Rev.\ D }{\bf #1},}
\def\PRL#1,{{\jnfont Phys.\ Rev.\ Lett.\ }{\bf #1},}
\def\MPLA#1,{{\jnfont Mod.\ Phys.\ Lett.\ A }{\bf #1},}
\def\JPG#1,{{\jnfont J.\ Phys.\ G}{\bf #1},}
\def\CTP#1,{{\jnfont Commun.\ Theor.\ Phys.\ }{\bf #1},}
\def\ZPC#1,{{\jnfont Z.\ Phys.\ C }{\bf #1},}
\def\JHEP#1,{{\jnfont JHEP \ }{\bf #1},}
\def\Rv{\not{\hbox{\kern-1pt $R$}}}
\def\p{\not{\hbox{\kern-3pt $p$}}}

\newcommand{\bea}{\begin{eqnarray}}
\newcommand{\eea}{\end{eqnarray}}

\newcommand{\bcen}{\begin{center}}
\newcommand{\ecen}{\end{center}}

\newcommand{\ee}{e^+e^-}

\newcommand{\beq}{\begin{eqnarray}}
\newcommand{\eeq}{\end{eqnarray}}

\def\t1{\tilde{t_1}}

\def\be{\begin{equation}}
\def\ee{\end{equation}}
\def\bea{\begin{array}}
\def\eea{\end{array}}
\def\beqa{\begin{eqnarray}}
\def\eeqa{\end{eqnarray}}
\def\beqas{\begin{eqnarray*}}
\def\eeqas{\end{eqnarray*}}

\def\bp{\begin{picture}}
\def\ep{\end{picture}}
\def\bc{\begin{center}}
\def\ec{\end{center}}
\def\bfig{\begin{figure}}
\def\efig{\end{figure}}

\def\bit{\begin{itemize}}
\def\eit{\end{itemize}}

\def\f{\frac}

\def\[{\left[}
\def\]{\right]}
\def\({\left(}
\def\){\right)}

\def\..{\left.}
\def\.{\right.}

\def\ep{\epsilon}

\begin{document}
\title{Dirac dark matter in $U(1)_{B-L}$ with Stueckelberg mechanism}

\author{Chengcheng Han$^1$, M.L. L$\rm \acute{o}$pez-Ib$\rm \acute{a}$$\rm \tilde{n}$ez$^{2}$,
Bo Peng$^{2,3}$, Jin Min Yang$^{2,3,4}$  }

\affiliation{
$^1$ School of Physics, KIAS, 85 Hoegiro, Seoul 02455, Republic of Korea \\
$^2$ CAS Key Laboratory of Theoretical Physics, Institute of Theoretical Physics, Chinese Academy of Sciences, Beijing 100190, P. R. China \\
$^3$ School of Physical Sciences, University of Chinese Academy of Sciences,  Beijing 100049, P. R. China \\
$^4$ Department of Physics, Tohoku University, Sendai 980-8578, Japan
}
\begin{abstract}
We investigate a $U(1)_{B-L}$ gauge extension of the Standard Model (SM) where the gauge boson mass
is generated by
the Stueckelberg mechanism.  Three right-handed neutrinos are added to cancel the gauge anomaly and hence
the neutrino masses can be explained. A new Dirac fermion could be a WIMP dark matter
whose interaction with the SM sector is mediated by the new gauge boson. Assuming the
perturbativity of the gauge coupling up to the Planck scale, we find that only the resonance region
is feasible for the dark matter abundance.  After applying the $\Delta N_{eff}$  constraints from the
current Planck experiment, the collider search constraints as well as the dark matter direct detection limits,
we observe that the $B-L$ charge of dark matter satisfies $|Q_{\chi}|>0.11$.
Such a scenario might be probed conclusively by the projected CMB-S4 experiment,
assuming the right-handed neutrinos are thermalized with the SM sector in the early universe.
\end{abstract}
\maketitle

\section{introduction}
The discovery of neutrino oscillations~\cite{Fukuda:1998mi,Ahmad:2002jz} indicates that neutrinos should have
tiny but non-zero masses, which can not be explained in the framework of the Standard Model (SM). One of the
compelling solutions to neutrino mass problem is to introduce three right-handed neutrinos which directly
couple to the SM sector through Yukawa interactions. At the same time, the gauge sector can be extended
with an additional anomaly free $U(1)_{B-L}$. Another fact demanding the presence of new physics is
the existence of dark matter (DM) which constitutes about 27\% of the global energy budget in the universe.
Therefore, it is intriguing to explain these phenomena in a same
framework \footnote{In Ref.~\cite{Krauss:2002px,Asaka2005,Ma:2006km,Boehm:2006mi,Kubo:2006yx,Aoki:2008av,Farzan:2010mr,Canetti:2012kh,Bonilla:2016diq,Batell:2017rol,Baumholzer:2018sfb,Blennow:2019fhy,FileviezPerez:2019cyn, Nanda:2019nqy, Cox:2017rgn, Kaneta:2016vkq},
a connection between the DM candidate and the origin of neutrino masses has been explored in detail.}.

In this article, we investigate a Dirac fermionic dark matter in a $B-L$ gauge extension
of the SM where the new gauge boson $Z'$
obtains mass via the Stueckelberg mechanism. Such a model is first proposed
by ~\cite{FileviezPerez:2019cyn} and here we will give a more comprehensive study.
In this model the neutrinos are Dirac fermions and a vector-like Dirac
particle $\chi$ charged under $U(1)_{B-L}$ is assumed to be a WIMP dark matter candidate.
In the early universe, DM is in thermal equilibrium with the SM plasma by exchanging the $Z'$ boson
and then freezes out when the expansion rate of the universe excesses its annihilation rate.
Finally, the current DM relic abundance needs to be consistent with the Planck data~\cite{Aghanim:2018eyx}.

On the other hand, since the right-handed neutrinos interact with the new gauge boson,  they are also in the
thermal equilibrium with the SM sector in the early universe. When the temperature goes much below the gauge
boson mass, they decouple and become the hot relic. Similar to the neutrinos and photons, they contribute to
the radiation energy density which is usually described as the effective number of neutrino species $N_{eff}$,
which is predicted to be 3.043 in the SM~\cite{Mangano:2005cc,Grohs:2015tfy,deSalas:2016ztq,Gariazzo:2019gyi,Bennett:2019ewm}.
The radiation density can be probed by the observation of the anisotropies in the cosmic microwave
background (CMB), which was proposed long time ago~\cite{Jungman:1995bz}.  The recent result from the
Planck satellite shows $N_{eff}=2.99_{-0.33}^{+0.34}$, providing a strong constraint on the extensions of SM
where the massless or light particles are present. We will show that this value already gives a very strong
limit on our scenario.
Particularly, it is pointed out in a recent article~\cite{Abazajian:2019oqj} that the projected CMB-S4
experiment will provide serious constraints for almost all Dirac-neutrino models, especially those
addressing the origin of small neutrino masses.

This paper is organized as follows. In Section~\ref{section2}, we introduce the $U(1)_{B-L}$ Stueckelberg
extension of the SM. Next, in Section ~\ref{section3}, the calculation of the shift in the effective number
of neutrino species, $\Delta N_{eff}$, and the current and future experimental bounds are discussed.
In Section ~\ref{section4}, we examine various constraints on the parameter space of the model.
At last, we draw our conclusions in Section~\ref{section5}.

\section{The Model}\label{section2}
The gauge group of the model is
\begin{equation}\label{gauge group}
SU(3)_C \bigotimes SU(2)_L \bigotimes U(1)_Y \bigotimes U(1)_{B-L}.
\end{equation}
The Stueckelberg mechanism as an alternative to the Higgs mechanism can give mass to abelian vector bosons
without breaking gauge invariance~\cite{Stueckelberg:1900zz, Kalb:1974yc, Allen:1990gb, Kors:2004dx, Kors:2004ri, Kors:2005uz, Ruegg:2003ps,Feldman:2011ms,Feldman:2006wb}. The Lagrangian related
to the Stueckelberg mechanism is given by
\begin{equation}\label{Lst}
{\cal L}_{St} = -\frac{1}{4} Z'^{\mu\nu} Z'_{\mu\nu} - \frac{1}{2} (M_{Z'}Z'_\mu + \partial_\mu \sigma)^2
\end{equation}
which is invariant under the following transformation
\begin{equation}\label{Trans}
\delta Z'_\mu = \partial_\mu \epsilon(x),\   \delta \sigma = -M_{Z'} \epsilon(x).
\end{equation}
In the quantum theory, a gauge fixing term
\begin{equation}\label{gauge fixing}
{\cal L}_{gf} = -\frac{1}{2\xi}(\partial_\mu Z'^\mu + \xi M_{Z'}\sigma)^2
\end{equation}
should be added to the total Lagrangian so that the new gauge boson becomes massive while the field $\sigma$ decouples. Note that the scalar field $\sigma$ can have Stueckelberg couplings to all abelian gauge bosons, including the hypercharge vector boson B in the SM~\cite{Feldman:2006wd,Feldman:2007wj,Abel:2008ai,Zhang:2009zzt,Burgess:2008ri}. However, in this work, we only focus on the pure Stueckelberg sector in the absence of the mass mixing of the gauge boson B with the $U(1)_{B-L}$ gauge boson $Z'$ for simplicity. Then the $B-L$ vector current $J'$
coupling to the gauge boson $Z'$  is given as
\begin{equation}\label{inreraction lagrangian}
{\cal L}^{int}_{St} = -g'Z'_\mu J'^\mu
\end{equation}
where $g'$ is the $B-L$ gauge coupling and $J'$ comes from quarks, leptons and DM.

In the Stueckelberg scenario, neutrinos are Dirac-type and their masses can be generated by Yukawa interactions via the Higgs mechanism,
\begin{equation}\label{neutrino yukawa}
{\cal L}_{\nu} = -Y_\nu \bar{l}_L i \sigma_2 H^* \nu_R + h.c.
\end{equation}
For a sub-eV neutrino mass, the coupling $Y_\nu$ should be generally smaller than $10^{-12}$. We can add a
Dirac fermion $\chi$ which only takes the $B-L$ charge and can be a dark matter candidate
(its stability can be guaranteed if its $B-L$ charge $Q_\chi$ is not equal to $\pm 1$, otherwise it will
mix with the right-handed neutrino and decay).  The relevant Lagrangian for the DM is then written as
\begin{equation}\label{L dark matter}
{\cal L}_{DM} = i\bar{\chi}\gamma^\mu D_\mu \chi - M_\chi \bar{\chi}\chi .
\end{equation}
The total Lagrangian in the model can be summarized as
\begin{equation}\label{total lagrangian}
{\cal L}_{tot} = {\cal L}_{SM} + {\cal L}_{St} + {\cal L}_{gf} + {\cal L}^{int}_{St} + {\cal L}_{\nu} + {\cal L}_{DM}.
\end{equation}

Based on the above Lagrangian, for $M_\chi \gg m_f$, DM can annihilate into
$e^+_i e^-_i,\bar{\nu}_i \nu_i,\bar{u}_i u_i, \bar{d}_i d_i$ ($i$ denotes three families of quarks
and leptons) via the gauge boson $Z'$.
The non-relativistic form for these annihilation cross sections is
\beqa\label{ann channel}
\sigma\upsilon(\chi\bar{\chi}\rightarrow f\bar{f}) &=& \frac{N_f^C Q^2_\chi Q^2_f g'^4}{2\pi}\sqrt{1-\frac{m^2_f}{M^2_\chi}}\frac{2M^2_\chi+m^2_f}{(4M^2_\chi-M^2_{Z'})^2+M^2_{Z'}\Gamma^2_{Z'}} \nonumber \\
\label{ann channel2}&\approx& \frac{N_f^C Q^2_\chi Q^2_f g'^4 M^2_\chi}{\pi[(4M^2_\chi-M^2_{Z'})^2+M^2_{Z'}\Gamma^2_{Z'}]}
\eeqa
where $\upsilon$ is the relative velocity of the annihilating DM pair,
$N_f^C$ is the number of colors of the final state SM fermions,
$Q_\chi$ and $Q_f$ represent the $B-L$ charges of DM and SM fermions,
and $\Gamma_{Z'}$ is the decay width of $Z'$ boson given by
\beqa\label{Z' decay width}
\Gamma_{Z'} &= &\sum_f \theta(M_{Z'}-2m_f)
\frac{N_f^C Q^2_f g'^2 M_{Z'}}{12\pi}\sqrt{1-\frac{4m^2_f}{M^2_{Z'}}}
\left(1+\frac{2m^2_f}{M^2_{Z'}}\right) \nonumber \\
&& + \theta(M_{Z'}-2m_\chi)\frac{ Q^2_\chi g'^2 M_{Z'}}{12\pi}\sqrt{1-\frac{4m^2_\chi}{M^2_{Z'}}}
\left(1+\frac{2m^2_\chi}{M^2_{Z'}}\right).
\eeqa
From Eq.(\ref{ann channel}) it can be seen that the ratio of the contribution of a quark to the total DM
annihilation cross section and the contribution of a lepton is about 1 : 3. Besides, DM is also able to annihilate into two $Z'$ bosons when DM is heavier than $Z'$. The annihilation cross section of this channel is
\beq\label{Z'Z'}
\sigma\upsilon(\chi\bar{\chi}\rightarrow Z' Z') = \f{Q^4_\chi g'^4}{16\pi M^2_\chi}\left(1-\f{M^2_{Z'}}{M^2_\chi}\right)^{3/2}\left(1-\f{M^2_{Z'}}{2M^2_\chi}\right)^{-2} .
\eeq

\section{The effective number of neutrino species in cosmology} \label{section3}
Three additional right-handed neutrinos can be in thermal equilibrium with the SM plasma via the exchange
of $Z'$ boson in the early universe so that they can contribute to the expansion rate of the universe.
However, due to their weak interactions\footnote{For a dark matter mass around TeV, the dark matter relic
abundance generally requires the gauge boson mass not much beyond that.}, such particles decouple earlier
from the plasma than the left-handed neutrinos and therefore their contribution to the energy density
of the universe is suppressed compared to that of the left-handed neutrinos. The extra radiation energy
density is usually expressed in terms of an effective number of neutrinos,
$N_{eff}=(8/7)(11/4)^{4/3}\rho_\nu/\rho_\gamma$. The SM prediction of this value is 3.043.

In this scenario, the extra contribution of the right-handed neutrinos to the effective number of
neutrino species is given as
\beq\label{deltaNeff}
\Delta N_{eff}=N_{\nu_R}\left(\f{T_{\nu_R}}{T_{\nu_L}}\right)^4=N_{\nu_R}\left(\f{g_{\ast}(T_{\nu_R}^{dec})}{g_{\ast}(T_{\nu_L}^{dec})}\right)^{-\f{4}{3}}
\eeq
where $N_{\nu_R}$ represents the number of relativistic right-handed neutrinos,
$g_{\ast}(T)=g_B(T)+\f{7}{8}g_F(T)$ with $g_{B,F}(T)$ being the number of bosonic and fermionic relativistic
degrees of freedom in equilibrium at the temperature T. The second equality is obtained from taking
into account of the isentropic heating of the rest of the plasma between $T_{\nu_R}^{dec}$ and $T_{\nu_L}^{dec}$
decoupling temperatures. Taking three active neutrinos, $e^{\pm}$ and photon into account, we have $g_{\ast}(T_{\nu_L}^{dec})=43/4$ at $T_{\nu_L}^{dec}\approx 2.3$ MeV~\cite{Enqvist:1991gx}.

The effective number of neutrino species has a strong relation with the temperature at which the right-handed neutrinos decouple from the SM plasma, which can be decided by
\beq
H(T_{\nu_R}^{dec}) \simeq \Gamma(T_{\nu_R}^{dec}).
\eeq
Here $H(T)$ is the Hubble expansion parameter which is estimated by $1.66 g^{1/2}_*(T) \frac{T^2}{M_{Pl}}$ where $g_*(T)$ is the effective degree of freedom~\cite{Bazavov:2011nk,Borsanyi:2016ksw} including the contribution of right-handed neutrinos.  $\Gamma_{\nu_R}(T)$ is the right-handed neutrino interaction rate which can be calculated by $ \sum_f n_{\nu_R}(T)\langle\sigma(\bar{\nu}_R \nu_R \rightarrow \bar{f}f)\upsilon\rangle$.

In our paper we examine the bounds on the parameters of the model by using present and prospective experimental data.
The current Planck CMB measurement gives the result $N_{eff}=2.99_{-0.33}^{+0.34}$
including baryon acoustic oscillation (BAO) data~\cite{Aghanim:2018eyx}.
Combining with the data given above, $N_{eff}^{SM}=3.043$, we adopt a conservative limit
$\Delta N_{eff} < 0.283$ and then the bound with respect to the $B-L$ gauge coupling $g'$
and $Z'$ boson mass $M_{Z'}$ is given as $M_{Z'}/g' > 10.4 ~{\rm TeV}$.
It gives a very strong limit on the parameter space, as we will show later.

Besides, there are several experiments with better sensitivities which are underway or projected.
The South Pole Telescope (SPT-3G), which is a ground-based telescope in operation at present,
will have a sensitivity of $\sigma(\Delta N_{eff})=0.058$~\cite{Benson:2014qhw}.
The CMB Simons Observatory (SO), which will see first light in 2021 and start a five-year survey in 2022,
is expected to reach a similar sensitivity in the range of $\sigma(\Delta N_{eff})=0.05 - 0.07$~\cite{Abitbol:2019nhf}.
The CMB Stage IV (CMB-S4) experiment will have the potential to constrain
$\Delta N_{eff}=0.06$ at $95\%$ C.L. as a single parameter extension to $\Lambda$CDM~\cite{Abazajian:2019eic}.
Importantly, according to Eq.(\ref{deltaNeff}), the minimal shift in the effective number of neutrino species
in our scenario can be evaluated to acquire $\Delta N_{eff} \geq 0.141$ when $T_{\nu_R}^{dec}$ is high enough. Hence, the future CMB-S4 experiment will be able
to probe this scenario for arbitrary decoupling temperatures conclusively as long as the right-handed
neutrinos have a thermalization with the SM plasma in the early universe.

\section{Numerical results and discussions}\label{section4}
According to the Lagrangian in Eq.(\ref{total lagrangian}),  in the dark matter sector there are only
four relevant parameters: the $U(1)_{B-L}$ gauge boson mass $M_{Z'}$, the DM mass $M_{\chi}$,
the $B-L$ charge of the DM $Q_{\chi}$ and the gauge coupling $g'$. By definition the $B-L$ charges of
quarks and leptons in SM are +1/3 and -1 respectively and DM could have arbitrary charge except
for $\pm 1$.  The gauge coupling should satisfy $g'<2\sqrt{\pi}$ to ensure the perturbativity of the theory.

In this work,  we consider the constraints from the DM relic density $\Omega_{\chi}h^2$,
the shift in the effective number of neutrino species $\Delta N_{eff}$, the dark matter direct detection limits
as well as the collider search limits for the $Z'$. We also require the gauge coupling to keep perturbativity
up to the Planck scale  $M_{pl}$.

The one-loop $\beta$ function of the $U(1)_{B-L}$ gauge coupling is given by
\beq\label{beta function}
\beta(g')=\f{{g'}^3}{16\pi^2}\sum_i Q_i^2 = \f{(60+9Q_{\chi}^2){g'}^3}{144\pi^2}  = \beta_0 {g'}^3
\eeq
where $i$ sums over all particles that carry $B-L$ charge and $\beta_0$ is a function of $Q_{\chi}$.
Assuming that the Landau pole does not occur below the Planck scale, then we get
\beq\label{landau pole}
g'(\mu) \lesssim \left(2\beta_0\ln\f{M_{pl}}{\mu}\right)^{-\f{1}{2}}
\eeq
with the renormalization scale $\mu=M_{Z'}$.

\begin{figure}[htp]
\centering
\includegraphics[width=3.2in,height=3in]{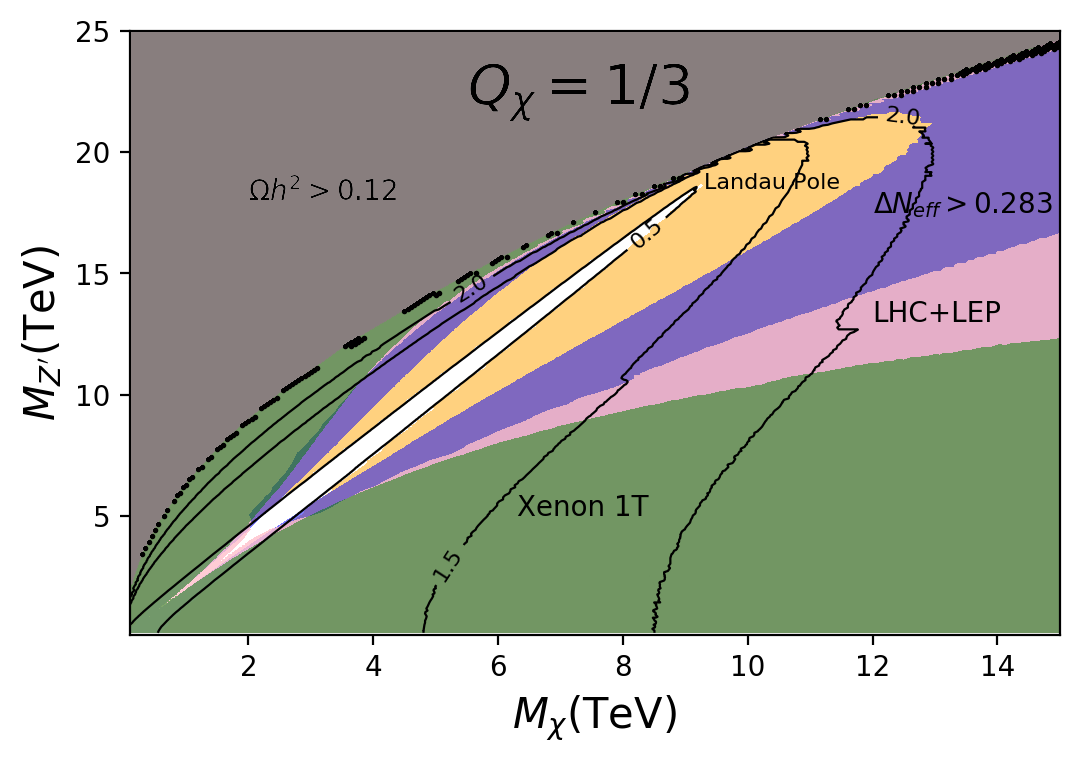}
\includegraphics[width=3.2in,height=3in]{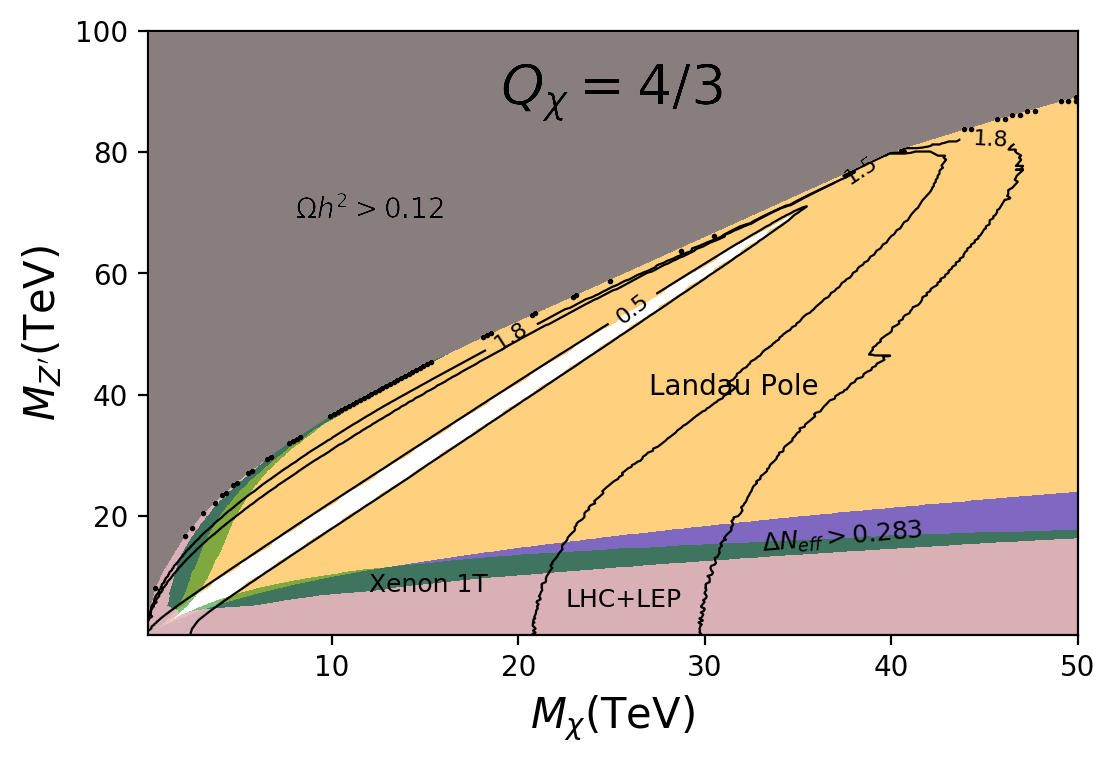}
\vspace*{-1.5cm}
\caption{Various constraints on the parameter space of $M_{\chi}$ versus $M_{Z'}$
for $Q_{\chi}=1/3$ (left) and $4/3$ (right). All the regions, except for the grey region
which overproduces DM, satisfy the DM relic density, $\Omega_{\chi} h^2 = 0.12$.
The solid black curve represents different $g'$s that are tuned to get the correct DM relic density.
The green, pink, blue and orange regions are exluded by the Xenon 1T~\cite{Aprile:2017iyp},
LHC+LEP~\cite{Aaboud:2017buh,Alioli:2017nzr}, $\Delta N_{eff}$~\cite{Aghanim:2018eyx} and Landau pole,
respectively.
Black points satisfy the DM relic abundance but their gauge couplings are larger than $2\sqrt{\pi}$.
The remained blank region survives all these constraints.}
\label{mxandmz}
\end{figure}

In our numerical calculation, we use LanHEP 3.2.0~\cite{Semenov:2014rea} to generate the Feynman rules
of the model and apply MicrOMEGAs 5.0.9~\cite{Belanger:2018mqt} to compute the DM relic abundance
and DM-nucleon scattering cross-section. In Fig.~\ref{mxandmz}, we show all the relevant constraints
on the plane of $M_{Z'}$ versus $M_{\chi}$ for $Q_{\chi}=1/3$ (left) and $4/3$ (right).
In the grey region in the upper left corner the dark matter is overabundant because the dark matter
annihilation cross section is too small. It also shows that the direct detections
and collider experiments can exclude some regions where the gauge boson mass or dark matter mass
is relatively light. However, $\Delta N_{eff}$ gives a stronger constraint than LHC+LEP and Xenon 1T.
For $Q_{\chi}=1/3$, the cosmological constraint can give an upper bound on $Z'$ boson mass and
DM mass: $M_{Z'}\leq 22$ TeV and $M_{\chi}\leq 13$ TeV,  which is consistent with the result in
Ref.~\cite{FileviezPerez:2019cyn}. For a larger $Q_{\chi}$, this constraint can be weakened because
a smaller $g'$ can still achieve the correct DM relic abundance. Assuming the gauge coupling
perturbative up to the Planck mass scale which generally requires
$g'  \lesssim 0.5$ for $Q_{\chi} =1/3$ or $4/3$, the survival region is restricted to
the blank area where the DM annihilation cross-section is enlarged significantly via the
resonance $M_{Z'} \simeq 2M_{\chi}$.  From this figure, we conclude that the gauge boson mass is
in general of tens of TeV after we impose the Landau pole constraint and hence the gauge boson
might be possibly accessible at the future 100 TeV hadron collider. Besides, the small bulge
which can be observed at about $M_{\chi}\approx 11.5(39)$  TeV in the left (right) panel
originates from the contribution of the $\chi \bar{\chi}\rightarrow Z' Z'$ channel to the relic density.

\begin{figure}[htp]
\centering
\includegraphics[width=4.2in,height=3.5in]{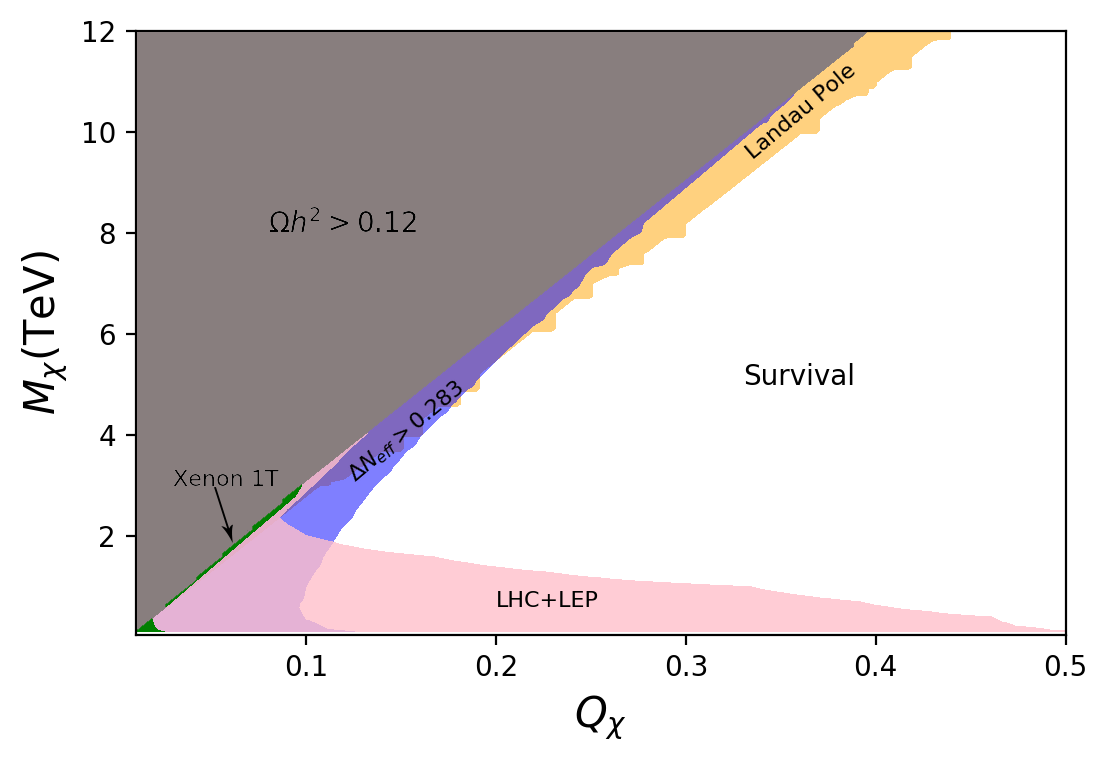}
\vspace{-.5cm}
\caption{Same as Fig.\ref{mxandmz}, but showing $M_{\chi}$ versus $Q_{\chi}$ in the resonant
region $M_{Z'}=2M_{\chi}$.}
\label{mxandnx}
\end{figure}

In Fig.\ref{mxandnx} we show the parameter space of $M_\chi$ versus $Q_\chi$ by setting $M_{Z'}=2M_{\chi}$
and changing $g^\prime$ to meet the condition $\Omega_{\chi} h^2 = 0.12$(note that we only scanned the positive $Q_{\chi}$). It shows that the Xenon-1T experiment
has a weaker constraint than others and collider experiments mainly restrict the lower bound on
DM mass while $\Delta N_{eff}$ and the Landau pole primarily restrict the upper bound.
As $Q_{\chi}$ increases, the restrictions of collider experiments and $\Delta N_{eff}$ get weakened
due to the decrease of $g'$. From this plot, we can see that the $B-L$ charge of DM must be larger
than 0.11 to evade all the constraints. For $Q_{\chi}=1/3$, $M_{\chi}$ should heavier than 1 TeV.
For the region $Q_{\chi}>0.5$, the current experimental data does not give limits on the parameter space
due to a small $g'$. Nevertheless,  the future CMB-S4 experiment could be able to cover almost
all the parameter space of this scenario even in the resonant region.

\section{Conclusion}\label{section5}
In this work, we studied dark matter in a $U(1)_{B-L}$ gauge extension of the Standard Model
where the $B-L$ gauge boson gains mass via the Stueckelberg mechanism.  Three right-handed neutrinos
are added to cancel the gauge anomaly and the neutrino masses can be thus explained.
A new Dirac fermion plays the role of WIMP dark matter while its interaction with Standard Model sector
is mediated by the new gauge boson. Assuming the perturbativity of the gauge coupling up to the Planck scale,
we found that only the resonance region is available for the dark matter abundance.  After applying
 the $\Delta N_{eff}$ constraints from the current Planck experiment, the collider search constraints
as well as the dark matter direct detection limits, we observed that the $B-L$ charge of dark matter
satisfies $|Q_{\chi}|>0.11$.  The projected CMB-S4 experiment might be able to probe this scenario
conclusively, assuming the right-handed neutrinos are thermalized with the Standard Model sector
in the early universe.

\section*{Acknowledgement}
This work was supported by the National Natural Science Foundation of China (NNSFC)
under grant Nos.11675242, 11821505, and 11851303, by Peng-Huan-Wu Theoretical
Physics Innovation Center (11947302), by the CAS Center for Excellence in Particle Physics
(CCEPP), by the CAS Key Research Program of Frontier Sciences and by a Key R\&D Program
of Ministry of Science and Technology under number 2017YFA0402204.

\end{document}